\documentclass[twocolumn,showpacs,preprintnumbers,amsmath,amssymb]{revtex4}

\usepackage{graphicx}
\usepackage{bm}

\begin{document}

\def\be{\begin{equation}}
\def\ee{\end{equation}}

\title{Adiabatic cooling of trapped nonneutral plasmas}
\author{Giovanni Manfredi}
\email{Manfredi@unistra.fr}
\author{Paul-Antoine Hervieux}
\affiliation{Institut de Physique et Chimie des Mat\'{e}riaux, CNRS
and Universit\'{e} de Strasbourg, BP 43, F-67034 Strasbourg, France}

\date{\today}
\begin{abstract}
Nonneutral plasmas can be trapped for long times by means of combined electric and magnetic fields. Adiabatic cooling is achieved by slowly decreasing the trapping frequency and letting the plasma occupy a larger volume. We develop a fully kinetic time-dependent theory of adiabatic cooling for plasmas trapped in a one-dimensional well. This approach is further extended to three dimensions and applied to the cooling of antiproton plasmas, showing excellent agreement with recent experiments [G. Gabrielse et al., Phys. Rev. Lett. {\bf 106}, 073002 (2011)].
\end{abstract}

\maketitle

{\it Introduction}.--- Nonneutral plasmas (consisting purely of electrons, ions, or positrons) have received much attention in the last few decades \cite{davidson}, particularly because very accurate table-top experiments can be devised to study their properties. Nonneutral plasmas can be confined for much longer periods of time (several days) compared to quasi-neutral plasmas and are more easily cooled down to low temperatures. Their confinement is usually achieved by means of a cylindrical device (Penning-Malmberg trap), in which an axial magnetic field ensures the radial confinement while the axial trapping is provided by a set of appropriately distributed electrodes.

The creation and confinement of nonneutral plasmas consisting of positrons or antiprotons ($\bar p$) is a major issue for the production of significant amounts of neutral antimatter in the form of antihydrogen ($\bar{H}$) atoms. Spectacular advances in this field have been achieved in recent years by several international collaborations, such as ALPHA \cite{alpha}, ATRAP \cite{atrap}, and ASACUSA \cite{asacusa}. The motivation of this recent and exciting research is to compare the physical properties of matter and antimatter, for instance the spectral lines of antihydrogen or its behavior under the influence of gravity. The recently established international collaboration GBAR aims at determining the gravitational acceleration of $\bar{H}$ atoms by letting them fall in the gravitational field of the Earth \cite{gbar}. For the same purpose, the AEGIS collaboration proposes to measure the shift of the interference pattern of free-falling antihydrogen atoms going through a series of gratings \cite{aegis}.

The $\bar{H}$ atoms are usually trapped in a magnetic well, where the effective confining potential has a depth of about 1K in temperature units \cite{alpha}. As the atoms are created with much higher energies, their cooling constitutes a major challenge to achieve an effective confinement.
The ALPHA collaboration \cite{alpha} recently reported confining several hundred  $\bar{H}$ atoms for more than 15 minutes. The antiprotons were first pre-cooled by interacting with cold electrons and then brought down to 40 K by evaporative cooling \cite{atrap2}. Unfortunately, during evaporative cooling most charged particles are lost, leading to a lower $\bar{p}$ density.
In contrast, using an adiabatic cooling technique, Gabrielse et al. \cite{gabrielse11} managed to cool down $3\times 10^6$ antiprotons to about 3K, with almost no losses observed.

Adiabatic cooling is achieved by lowering the trapping frequency $\omega$ that confines a nonneutral plasma within a one-dimensional (1D) harmonic well, $U(z)=m\omega^2(t)z^2/2$. If the plasma density is low enough, one can adopt a single-particle approach. For a slowly varying frequency, the action $E/\omega$ is conserved, so that the energy (and the temperature) of the plasma decreases linearly with the frequency, i.e., $T \sim \omega$.
When the plasma density is high enough, the effect of the space charge cannot be neglected and the single-particle approach breaks down. Li et al. \cite{GZLi} have developed a thermodynamic theory of adiabatic cooling for an ion cloud in a Penning trap, assuming that the ions always remain at Maxwell-Boltzmann equilibrium \cite{note1}.

The purpose of this Letter is to develop a fully kinetic description of the adiabatic expansion and cooling of a nonneutral plasma, without making any assumptions on its thermodynamic properties except for the initial equilibrium.
The results will be compared to recent experiments from the ATRAP collaboration \cite{gabrielse11}.

{\it Kinetic model in 1D}.---
In most experiments, the charged particles are confined in a harmonic potential well, with a strong magnetic field along the axial (parallel) direction $z$, which ensures radial (perpendicular) confinement. It is then reasonable to restrict our analysis to a 1D geometry along the $z$ axis.
Nevertheless, particle collisions can couple the axial and radial motion, leading to energy equipartition between the parallel and perpendicular temperatures. Thus, from a thermodynamic point of view, the expansion is 1D when collisions are rare, and 3D when collisions are dominant \cite{dubin}.

We first concentrate on the collisionless 1D regime, where the antiproton plasma can be described by the Vlasov-Poisson equations:
\begin{eqnarray}
\frac{\partial{f}}{\partial{t}} &+&
v_z\frac{\partial{f}}{\partial{z}} -\left(\omega^2(t) z +\frac{eE_z}{m}\right) \frac{\partial{f}}{\partial{v_z}} = 0, \label{vlasov} \\
\frac{\partial E_z}{\partial z} &=& -
\frac{e \sigma_\perp}{\varepsilon_0} \int f(z,v_z,t)~dv_z ,
\label{poisson}
\end{eqnarray}
where $-e$ and $m$ are the antiproton charge and mass, $E_z$ is the parallel self-consistent electric field, $\sigma_\perp$ is the 2D number density in the perpendicular plane, and $n=\int fdv_z$ is the 1D density along $z$.
The initial condition is computed self-consistently by assuming that the plasma is at thermal equilibrium with temperature $T_0$.

If the frequency $\omega$ decreases in time, the plasma will experience an expansion. In order to follow this expansion, it is useful to use the scaling techniques introduced in Refs. \cite{manfredi93,burgan}. We define the new phase space variables $(\xi, \eta)$ and the new time $\theta$ in the following way:
\be
z=C(t)\xi;~~ dt=A^2(t)d\theta;~~
v_z=(C/A^2)\eta + \dot{C} \xi. \label{scaling}
\ee
Our aim is to find the ``right" scaling transformations such that the plasma will be frozen in the scaled variables. Then, the scaling factor $C(t)$ will inform us on the expansion law, which in turn can be related to the evolution of the plasma temperature.

We also need to transform the distribution function in order to guarantee the invariance of the total number of particles, $f(z,v_z,t)dzdv_z=F(\xi,\eta,\theta)d\xi d\eta$. Finally, the scaled electric field is defined as $\mathcal{E}(\xi,\theta) = E_z(z,t)$ in order to keep Poisson's equation unchanged: $\partial_\xi \mathcal{E} = -\frac{e\sigma_\perp}{\varepsilon_0}\int F d\eta$.
In the scaled variables, the Vlasov equation reads as follows:
\be
\frac{\partial{F}}{\partial \theta} + \eta\frac{\partial{F}}{\partial \xi}+ \frac{\partial}{\partial \eta} \left[Q(\xi,\eta,\theta) F(\xi,\eta,\theta) \right]= 0~,
\label{vlasov-scaled}
\ee
where the scaled acceleration is
\be
Q=-\frac{A^4 e\mathcal{E}}{mC} - \left( \frac{A^4\ddot{C}}{C} +\omega^2A^4\right)\xi + 2A^2 \left(\frac{\dot{A}}{A}-  \frac{\dot{C}}{C} \right)\eta. \label{scaled-force}
\ee

Let us now introduce some specific expressions for the scaling coefficients:
$A(t)=(1+\Omega t)^\alpha$ and $C(t)=(1+\Omega t)^\delta$. In addition, we postulate a similar form for the time-dependent trapping frequency: $\omega(t)=\omega_0(1+\Omega t)^{-\beta}$, where $\alpha$, $\beta$, and $\delta$ are real positive numbers. Here, $\Omega>0$ is a parameter that determines the rapidity of the change of the trapping frequency. With these prescriptions, the scaled acceleration becomes:
\begin{eqnarray}
Q &=& -(1+\Omega t)^{4\alpha-\delta}e\mathcal{E}/m \nonumber \\
 &-&[\delta(\delta-1)\Omega^2(1+\Omega t)^{4\alpha-2}
+\omega_0^2(1+\Omega t)^{4\alpha-2\beta}]\xi \nonumber \\
&+& 2\Omega(\alpha-\delta)(1+\Omega t)^{2\alpha-1} \eta \label{Qgeneral}.
\end{eqnarray}
The idea is to choose the parameters $\alpha$ and $\delta$ so that as many terms as possible in Eq. (\ref{Qgeneral}) become time-independent. For all values of $\beta$, we need to choose $\alpha=\delta/4$ so that the coefficient of the first term is constant. Then, two cases clearly appear. For $\beta \ge 1$ one should take $\delta=2$,
whereas for $\beta < 1$ one should have $\delta=2\beta$.
Note that for $\beta=1$ all coefficients become constant.

The transformed equations describe a ``virtual" physical system made of charged particles interacting through (i) the self-consistent electric field $\mathcal{E}$, (ii) a harmonic field composed of a constant and a decreasing-in-time term, and (iii) a friction field, either constant or decreasing with time. In the scaled phase space, such a system will relax to a steady state with zero velocity (because of the friction) and uniform spatial density $n_0$ (defined by $e^2 \sigma_\perp n_0/m\varepsilon_0=\omega_0^2$), thus creating a linear repulsive force $\mathcal{E}$ that exactly cancels the second term of Eq. (\ref{Qgeneral}).

{\bf
The approach to the asymptotic solution is nicely seen in Fig. 1, where we plot the initial and final densities. The asymptotic profile is dictated by the structure of the scaled equations (\ref{vlasov-scaled})-(\ref{Qgeneral}) and thus constitutes a {\em universal attractor} towards which any initial condition (not necessarily an equilibrium) should converge. This is true even for rapid variations of $\omega(t)$, thus generalizing results that rely on assumptions of thermal equilibrium and adiabatic expansion \cite{GZLi}.
}
\begin{figure}[htb]
\includegraphics[height=4.cm]{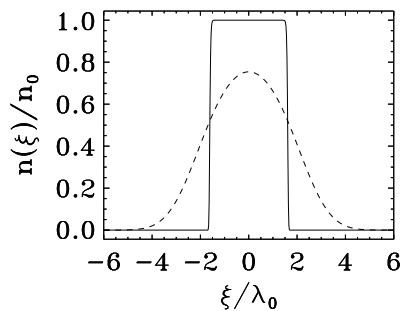}
\caption{Density profiles in the scaled space at the beginning (dashed line) and at the end (solid line) of the simulation.}
\end{figure}

As the system becomes frozen in the $\xi$-space, the expansion law is determined by the scaling function $C(t)$ and hence the exponent $\delta$: $z \sim t^{2\beta}$ for $\beta\le 1$ and $z \sim t^{2}$ for $\beta > 1$.
It is easy to understand why the value $\beta=1$ plays a pivotal role: the expansion is adiabatic when $|\dot{\omega}/\omega|\ll \omega$, which implies $\beta \Omega\ll \omega_0(1+\Omega t)^{1-\beta}$. When $\beta>1$, the previous expression cannot be satisfied for all times. In contrast, for $\beta\le 1$, the expansions is always adiabatic, provided that $\beta\Omega<\omega_0$. In the following we will concentrate on such adiabatic regimes.

In the scaled space, the density approaches asymptotically the value $n(\xi)=n_0$ \cite{note2}. Using the conservation of the total number of particles $n(z)dz=n(\xi)d\xi$ to transform back to real variables, one finds that the density behaves as $n(z,t)=n_0(1+\Omega t)^{-2\beta}$. Finally, using the isentropic law for a weakly coupled plasma $T \sim n^{\gamma-1}$ [with $\gamma=(d+2)/d=3$ in 1D] and the expression for $\omega(t)$, one obtains the temperature as a function of the frequency: $T/T_0 = (\omega/\omega_0)^{4}$.

An important consequence is that the plasma frequency behaves as: $\omega_p(t)=(e^2 \sigma_\perp n(t)/m\varepsilon_0)^{1/2}=\omega_0/(1+\Omega t)^{\beta}\equiv \omega(t)$. In other words, the plasma frequency approaches the harmonic frequency and remains locked to it for the rest of the evolution.

The relation $T\sim \omega^4$ represents an asymptotic law. However, some transient regime must exist in the case of low-density plasmas, for which the Coulomb interaction is almost negligible. In this case, the expansion should initially follow the single-particle law $T \sim \omega$. In contrast, for high-density plasmas the asymptotic law should be observed from the beginning of the evolution.
These two regimes can be characterized by the total number of particles in 1D, $N=\int f dz dv_z$ (normalized to $n_0 \lambda_0$, where $\lambda_0=\sqrt{k_B T_0/m}/\omega_0$).

In order to check the above conjectures, we performed numerical simulations of the nonlinear Vlasov-Poisson equations (\ref{vlasov})-(\ref{poisson}), starting from self-consistent equilibria characterized by different values of $N$. The equations were solved in the scaled phase space $(\xi,\eta)$ using an Eulerian Vlasov code \cite{filbet}.

In Fig. 2, we plot the temperature at the center of the plasma cloud ($z=0$) against the inverse harmonic frequency, for $\beta=1$, $\Omega=0.02\omega_0$, and three values of $N$. As expected, for $N<1$ the temperature evolves first as $\omega$ (single-particle transient) and later as $\omega^4$ (self-consistent asymptotic behavior). For $N>1$, the asymptotic behavior is observed right from the start. The same pattern was observed for runs with $\beta<1$.
\begin{figure}[htb]
\includegraphics[height=4.cm]{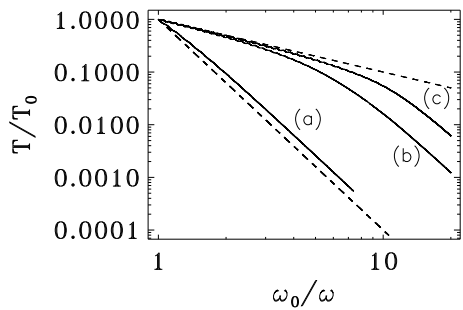}
\caption{Plasma temperature as a function of the inverse of the harmonic frequency $\omega$ in the 1D regime, for $N=10$ (a), $N=0.3$ (b), and $N=0.1$ (c). The dashed straight lines have slopes $-1$ and $-4$.}
\end{figure}

{\it Extension to 3D}.---
The above model, Eqs. (\ref{vlasov})-(\ref{poisson}), is a mean-field approximation that neglects binary collisions between particles. However, collisions are important in antiproton cooling experiments and their main effect is to drive the system towards energy equipartition between the parallel and perpendicular motion. In order to model collisions, we add a relaxation term on the right-hand side of Eq. (\ref{vlasov}):
\be
\left(\frac{\partial{f}}{\partial{t}}\right)_{\rm coll} = -\nu_{\rm coll}\left( f-\frac{n e^{-m (v_z-u)^2/2 k_B T_\perp}}{(2\pi k_B T_\perp/m)^{1/2}} \right),
\label{relax}
\ee
where $u(z,t)=\int f v_z dv_z/n$, $T_\perp(t)$ is the perpendicular temperature, which obeys the following equation:
\be
\frac{dT_\perp}{dt} =  -\frac{\nu_{\rm coll}}{2}(T_\perp -T_\parallel),
\label{Tperp}
\ee
and $k_B T_\parallel(t)=m\int f(0,v_z,t) v_z^2 dv_z/n(0,t)$ is the parallel temperature at the center of the plasma. A factor 1/2 appears in Eq. (\ref{Tperp}), but not in Eq. (\ref{relax}), because the perpendicular motion is associated with two degrees of freedom.

In 1D, the temperature depends on the frequency as $T \sim \omega^{4}$ and the density as $n \sim  \omega^2$, implying that $\omega_p(t) \sim \omega(t)$.
This relation does not depend on the dimensionality of the system and can be extended to the 3D case. Using the isentropic law in three dimensions, $T \sim n^{2/3}$, yields the expression $T \sim \omega^{4/3}$. This is the law that we expect to observe in the collisional 3D regime.

In order to check this conjecture, we solve Eqs. (\ref{vlasov}), (\ref{poisson}), (\ref{relax}) and (\ref{Tperp}) using the same scaling techniques introduced above. Notice that in Eq. (\ref{Qgeneral}) we have two free parameters, namely $\alpha$ and $\gamma$. The latter determines the expansion of the plasma density and should be kept the same as in the 1D simulations, i.e. $\gamma=2\beta$ for the adiabatic regime. The exponent $\alpha$ can be chosen so that, in the scaled variables, the distribution function is bounded in velocity space. The relationship between the temperature in the real and scaled spaces is as follows: $\widehat{T}(\xi,\theta)=(A^4/C^2)T(z,t)$. If $T \sim \omega^{4/3}$, then, in order for the scaled temperature $\widehat{T}$ to remain constant, we need to take $\alpha=4\beta/3$. In addition, by choosing $\beta=3/8$ several terms in Eq. (\ref{Qgeneral}) are algebraically simpler, for instance the friction term becomes time-independent.
With these parameters, the distribution function should remain bounded in a finite phase-space volume in the scaled variables, thus greatly simplifying the numerical integration without any loss of generality as to the physics. Equations (\ref{relax})-(\ref{Tperp}) are also scaled using the same technique.

The numerical results confirm our conjecture. The case depicted in Fig. 3 represents a simulation with $\Omega=0.02 \omega_0$, $N=30$, and $\nu_{\rm coll}=0.001 \omega_0$. Initially, when the collision term has not yet had time to act, the parallel temperature follows the 1D law $T_\parallel \sim \omega^4$, while $T_\perp$ remans almost constant. On a longer timescale, however, the collisions drive the system towards energy equipartition, and both temperatures start decreasing as $\omega^{4/3}$. For larger collision rates, the law $T \sim \omega^{4/3}$ is observed from the beginning of the simulation. The total temperature is defined as $T={1 \over 3} T_\parallel + {2 \over 3} T_\perp$.

\begin{figure}[htb]
\includegraphics[height=4.5cm]{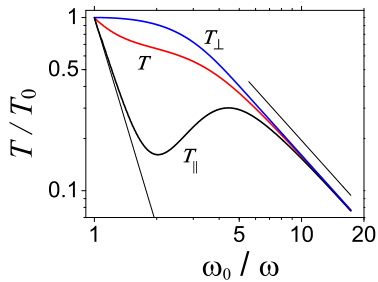}
\caption{(Color online). Parallel, perpendicular, and total temperatures as a function of the inverse of the instantaneous trapping frequency $\omega(t)$ in the 3D regime. The straight lines have slopes $-4$ and $-4/3$.}
\end{figure}

{\it Comparison to experiments}.---
We now have all the elements to perform a detailed comparison with the recent experimental results of Ref. \cite{gabrielse11}, where $N_{\bar{p}}=5\times 10^5$ antiprotons were confined in a cylindrical trap of volume $V=\pi R^2 L_z$, with radius $R \approx 2$ mm and length $L_z\approx 10$ mm. The axial magnetic field was $B=3.7~\rm T$. Each point on Fig. 3 of Ref. \cite{gabrielse11} corresponds to a different experiment in which the harmonic trap frequency is lowered from $\omega_0/2\pi$ (which ranges bewteen 90 kHz and 3 MHz) to the final value $\omega/2\pi=75$ kHz. During the ensuing expansion, the antiproton plasma cools down from $T_0=31~\rm K$ to a final temperature $T$. The total duration of the cooling is $t_f = 100 ~\rm ms$.

The experiments show that the final temperature behaves as $T \sim \omega^{1.2}$, which is rather close to the law $T \sim \omega^{4/3}$ that arises from our 3D model. In addition, the final temperature levels off at a value $T \approx 3.5~\rm K$ and does not decrease any further for frequencies above $\omega_0/2\pi = 500~\rm kHz$. This unexpected saturation is still unexplained and was tentatively attributed by the authors to intrinsic limits on the measuring apparatus, to some source of technical noise, or to some unspecified new physics appearing at high frequencies.

In order to perform our simulations, we need to specify three dimensionless parameters, namely $N$, $\Omega/\omega_0$, and $\nu_{\rm coll}/\omega_0$.
In 3D, $N=\int fdzdv_z$ is related to the total number of antiprotons by the expression: $N = N_{\bar{p}}/(n_0\lambda_0 \sigma_\perp \pi R^2)$. The parameter $\Omega$, which determines the rapidity with which the harmonic frequency decreases, can be expressed in terms of the cooling time $t_f$ by using the definition of $\omega(t)$.

Finally, the collisional equipartition rate was determined using the detailed calculations of Ref. \cite{glinsky}. For an unmagnetized plasma, the rate is roughly $\nu_{\rm unmag}=\sqrt{2k_BT/m} (N_{\bar{p}}/V) b$, where $b=e^2/(2\pi \varepsilon_0 k_B T)$, which yields $\nu_{\rm unmag} \approx 3300~\rm s^{-1}$ for the experimental parameters of interest here. The effect of the magnetic field is to reduce substantially the collision frequency. For $B=3.7 ~\rm T$, the correction factor can be estimated to be around 0.2 \cite{glinsky}, yielding $\nu_{\rm coll} \approx 660 ~ \rm s^{-1}$.

\begin{figure}[htb]
\includegraphics[height=5.2cm]{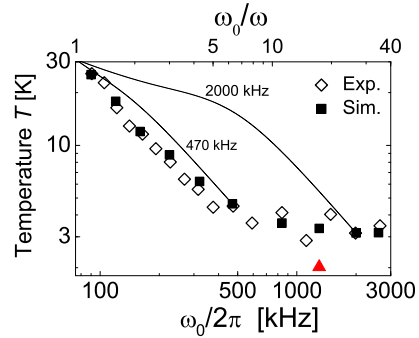}
\caption{(Color online). Final temperature $T$ after adiabatic cooling against the initial harmonic frequency $\omega_0/2\pi$. Black squares: present simulations; open diamonds: experimental results from \cite{gabrielse11}. The red triangle corresponds to a case where $\nu_{\rm coll}$ has been increased by a factor 5. The thin solid lines show the temperature evolution as a function of $\omega_0/\omega$ (top scale) for two values of $\omega_0$.}
\end{figure}

To compare with the experimental results, we performed a set of runs for different values of $\omega_0$, which correspond to different dimensionless parameters.
Two such runs are displayed as solid lines on Fig. 4.
In the same figure, we show the final temperature after adiabatic cooling as a function of the initial trapping frequency. The agreement between the simulations and the experimental results from Ref. \cite{gabrielse11} is excellent. Not only is the slope of the temperature curve $T(\omega_0)$ correct for small values of the harmonic frequency, but the leveling off at high frequencies is also reproduced with great accuracy.

Physically, this saturation is linked to the  strength of the collision rate relative to $\omega_0$. On the left of the diagram, $\nu_{\rm coll}/\omega_0$ is relatively large, so that equipartition occurs early and the corresponding points are roughly aligned along a line with slope $4/3$ (see the curve for $\omega_0/2\pi = 470$ kHz on Fig. 4). On the right of the diagram, $\nu_{\rm coll}/\omega_0$ is smaller and the temperature undergoes a transient with shallower slope (see curve for $\omega_0/2\pi = 2$ MHz on Fig. 4 and also the total temperature curve on Fig. 3), during which the parallel and perpendicular temperatures have not yet equilibrated. This results
in a final total temperature that is higher than what could be expected from the law $T \sim \omega^{4/3}$.
Indeed, by artificially increasing $\nu_{\rm coll}$ by a factor 5, we could lower the final temperature to roughly 2 K (red triangle in Fig. 4). Since the collision rate is very sensitive to the external magnetic field,
this suggests that lower temperatures may be achieved by reducing the latter, a conjecture that could easily be tested experimentally.

In summary, we have presented a fully kinetic theory for the adiabatic cooling of charged particles both in 1D and 3D. Our model, which does not make any explicit thermodynamical assumptions nor requires any fitting parameters, was capable of reproducing with great accuracy the results of recent experiments of the ATRAP collaboration \cite{gabrielse11}.

We thank Dr. W. Oelert and Dr. P. Perez for providing information on the experiments. This work was partially funded by the Agence Nationale de la Recherche, contract ANR-10-BLAN-0420.


\begin{references}
\bibitem{note1} This is somewhat in contradiction with their assumption of weak coupling.
\bibitem{note2} For $\beta=1$ the asymptotic density is $n(\xi)/n_0=(1+2\Omega^2/\omega_0^2)$, but we always assume $\Omega \ll \omega_0$.
\end{references}
\end{document}